\documentstyle{elsart}
 
\def\Journal#1#2#3#4{{#1} {\bf #2}, #3 (#4)}

\def\APP{{\em Acta Phys. Pol.} B}

\def\NPA{{\em Nucl. Phys.} A}

\def\PLB{{\em Phys. Lett.}  B}
\def\PRL{\em Phys. Rev. Lett.}
\def\PR{\em Phys. Rev.} 
\def\PREP{\em Phys. Rep.} 
\def\PRC{{\em Phys. Rev.} C}

\begin{document}
\hspace{9.8 cm}FZJ--IKP(TH)--1998--19
\begin{frontmatter}
 
\title{Role of the $\Delta$ isobar in the reaction 
$NN \rightarrow NN\pi$ near threshold}
 
\author{
C. Hanhart$^{a,b}$, J. Haidenbauer$^b$, O. Krehl$^b$ and J. Speth$^b$}

{\small $^a$Institut f\"ur Theoretische Kernphysik, Universit\"at Bonn,}\\{\small  D--53115 Bonn, Germany} \\
{\small $^b$Institut f\"{u}r Kernphysik, Forschungszentrum J\"{u}lich
GmbH,}\\ {\small D--52425 J\"{u}lich, Germany} \\
 

\begin{abstract}
A model calculation for pion production in nucleon--nucleon
collisions is presented. Direct production, pion rescattering and
contributions from pair diagrams are taken into account. The amplitudes
for the elementary processes are
based on well established microscopic models of nucleon--nucleon and 
pion--nucleon scattering. The $\Delta (1232)$ is included 
explicitly and is found to play an important role even at energies 
close to production threshold. A good overall agreement with
existing data from the pion production threshold up to the Delta
resonance region is achieved. 
\end{abstract}
\end{frontmatter}
 
Over the last decade or so a wealth of rather accurate data on pion
production in nucleon--nucleon ($NN$) collisions near threshold has
become available. This concerns reaction channels with a two--body
final state such as $pp\rightarrow d\pi^+$  \cite{TR2,CO,IU3}, as well as channels with
three--body final states such as $pp\rightarrow pp\pi^0$ \cite{IU1,TR1,UP} or 
$pp\rightarrow pn\pi^+$ \cite{Da,Ha}. The same period of time also witnessed 
lively activitiy on the theoretical side. In this case, however, most
of the efforts concentrated on the study of one particular reaction, 
namely the process $pp\rightarrow pp\pi^0$. Clearly this development 
was initiated by the unexpected failure of the first model studies 
\cite{MuS,NIS1}
to reproduce the corresponding data \cite{IU1}. The underprediction
of the $pp\rightarrow pp\pi^0$ total cross section by a factor of 5
turned out to be a challenging problem for theorists. 

These first investigations followed very closely the approach of
Koltun and Reitan \cite{KuR}. In this model it is assumed that
the pions are produced either directly from the nucleon 
(Fig.~\ref{beitraege}a) or via pion rescattering
(Fig.~\ref{beitraege}b), where in the latter the $\pi N$ amplitude is
approximated by the corresponding $\pi N$ s--wave scattering lengths. 
Subsequently two "new" mechanisms have been suggested that, in principle,
allow one to obtain quantitative agreement with the data on $\pi^0$ production
close to threshold. In one of them the pion is produced via an intermediate
virtual antinucleon--nucleon state in conjunction with the exchange of
a heavy meson ($\rho$, $\omega$, $\sigma$, ...)(HME), as depicted in
Fig.~\ref{beitraege}c. It was found in Refs. \cite{LuR,HMG} that the contributions 
from $\sigma$ and $\omega$ exchange can enhance the production cross section 
significantly enough to reproduce the data. 
In the other mechanism the static $\pi N$ interaction (i.e.,
the on--shell $\pi N$ amplitude at threshold) used in the Koltun--Reitan 
model for the rescattering
process is replaced by an off--shell $\pi N$ amplitude. Physically, the 
latter is required anyway because the exchanged pion in the rescattering
process is off--mass--shell. Since the isoscalar part of the $\pi N$
amplitude, which is responsible for rescattering in the $\pi^0$ production, 
is practically zero on--shell and at threshold due to constraints of
chiral symmetry, but can be fairly large once one goes off--shell, the use
of an off--shell amplitude, given by a certain model, leads likewise to an appreciable enhancement of
the production cross section \cite{HO,pap1,pap2}. 

Despite these apparent successes there is still a controversy over
which of the two processes if any, is the physically correct one. 
The HME mechanism has been put into question in a recent paper by 
J. Adam et al. \cite{Adam}. These authors work in the context of the 
relativistic Gross equation where contributions from intermediate
antinucleon--nucleon states are generated automatically by the scattering
equation and summed up to all orders. They see only moderate effects from 
these virtual intermediate states, resulting in contributions that are 
considerably smaller than the ones obtained in the perturbative treatment
of Refs. \cite{LuR,HMG}. 
Likewise, results based on the off--shell $\pi N$
amplitude have come into dispute. 
Calculations done in lowest order chiral perturbation theory \cite{CP1,CP2,CP3}
give rise to a $\pi N$ amplitude that differs in sign from the one employed 
in Refs. \cite{HO,pap1,pap2} when extrapolated off--shell. Accordingly, 
the amplitude from the rescattering diagram also changes sign and
then interferes destructively with the contributions from the other
production mechanisms. Consequently, agreement with the data in this case
can be no longer achieved. For a thorough discussion of this topic see
also Ref. \cite{HHH}. Furthermore we want to call attention to a recent paper 
by Bernard et al. \cite{ulfneu}, where it is argued that heavy baryon 
chiral perturbation theory (on which the works \cite{CP1,CP2,CP3} are based)
can not be used in reactions with as large momentum transfers as they are
typical for meson production in nucleon--nucleon collisions.

One possible way to learn more about the role of the individual production
mechanisms has been suggested in Ref. \cite{pap2,HHH}.
It consists of a comparative study of all relevant pion production channels
($pp \rightarrow pp\pi^0$, $pp \rightarrow pn\pi^+$, 
$pp \rightarrow d\pi^+$ and $pn \rightarrow pp\pi^-$) in a consistent approach (i.e., with the same
model and same parameters). Moreover, higher partial waves should be
taken into account. Most of the aforementioned investigations consider
only the lowest partial waves in the outgoing channel, which means that
the $NN$ system is in an S--wave state (or in the deuteron bound state) and
the pion is likewise in an s--wave relative to the nucleon pair. Such
calculations permit only conclusions on the absolute magnitude of the
production cross section near threshold. The inclusion of higher partial
waves, in the $NN$ as well as the $\pi N$ sector, allows one to calculate
predictions for differential cross sections and, in particular,
spin--dependent observables.
Therefore it is possible to examine whether the considered production 
mechanisms lead to the proper onset of higher partial waves, as 
suggested by the data. 
Results of the latter kind are particularly interesting because they 
reflect the spin--dependence of the production processes
and therefore should be very useful in discriminating between different
mechanisms. 

In this letter we present a model calculation for the reactions 
$pp \rightarrow pp\pi^0$, $pp \rightarrow pn\pi^+$,
$pp \rightarrow d\pi^+$, and $pn \rightarrow pp\pi^-$. 
It is an extension of our earlier study for s--wave pion production 
\cite{pap1,pap2} to higher partial waves. 
Now all $NN$- and $\pi N$ partial waves up to orbital angular momenta
$L = 2$, and all states with relative orbital angular momentum 
$l \leq 2$ between the $NN$
system and the pion are considered in the final state. 
Furthermore, the excitation of the $\Delta (1232)$
resonance is taken into account explicitly. Therefore we are able to
calculate meaningful predictions for the different reaction channels of  
$NN \to  NN\pi$ from their threshold up to the $\Delta$ resonance region. 

The reaction $NN \to  NN\pi$ is treated in a distorted wave born approximation,
in the standard fashion. The actual calculations are carried out 
in momentum space. For the distortions in the initial and final $NN$ states
we employ the model CCF of Ref. \cite{HHJ}. This potential has been 
derived from the full Bonn model \cite{MHE87}
by means of the folded--diagram expansion. 
It is a coupled channel ($NN$, $N\Delta$, $\Delta\Delta$) model that
treats the nucleon and the $\Delta$ degrees of freedom on equal footing.
Thus, the $NN \leftrightarrow N\Delta$ T--matrices that enter in the 
evaluation of the pion production diagrams involving the $\Delta$ isobar 
(cf. Fig. \ref{beitraegeND}) and the $NN$ T--matrices that are used 
for the diagrams in Fig. \ref{beitraege} are consistent solutions of 
the same (coupled--channel) Lippmann--Schwinger--like equation. 

The $\pi N \to \pi N$ T--matrix needed for the rescattering process is
taken from a microscopic meson--exchange model developed by the J\"ulich 
group \cite{SHH}. This interaction model is based on the conventional (direct 
and crossed) pole diagrams involving the nucleon and $\Delta$ isobar as 
well as t--channel meson exchanges in the scalar ($\sigma$) and vector
($\rho$) channel derived from correlated $2\pi$--exchange.  
Note that in our model of the reaction $NN\rightarrow NN\pi$
contributions where the pions are produced directly from the
nucleon or $\Delta$ (cf. Figs.\ref{beitraege}a and \ref{beitraegeND}a--c) 
are taken into account
explicitly. Therefore, the corresponding nucleon and $\Delta$ pole terms 
have to be taken out of the $\pi N$ T--matrix in order to avoid double 
counting. 

For the $\pi NN$ and $\pi N\Delta$ coupling constants at the pion 
production vertices we take the values $f_{NN\pi}^2/4\pi$ = 0.0778 \cite{HHJ} 
and  $f_{N\Delta\pi}^2/4\pi$ = 0.26 \cite{hoehler}. 
%
%
The form factors at 
these vertices are chosen to be soft (We use a monopole form with a cutoff 
mass $\Lambda _\pi$ = 900 MeV) in line with recent QCD lattice calculations
\cite{FoFa} and other information \cite{CoSc,ffulf}. 
The width of the $\Delta$ isobar is taken into account by using a 
complex $\Delta$ energy in the propagator. Specifically, we employ a 
parameterization of the width given by Kloet and Tjon in Ref. {\cite{KlTj} 
which is energy-- as well as momentum--dependent. We wish to point out,
however, that the width influences the observables only for 
energies $T_{lab} \geq$ 420 MeV.  For lower energies the results
obtained with and without $\Delta$ width are practically the same. 
Note that for simplicity we have suppressed 
the three--body singularity that appears in the pion rescattering 
diagram for energies above threshold by fixing the corresponding
three--body propagator to its threshold value. 
Thus, three particle unitarity is not fulfilled in our calculation.  
Earlier studies \cite{charlotte} have shown, however, that a large part of 
the imaginary part produced by the three particle cut of the pion exchange 
is cancelled by the imaginary part arising from the nucleon self energy 
contribution above pion threshold.
Therefore we expect the effect of our approximation to be small, at 
least for energies close to threshold.  
Furthermore, a 
non--relativistic boost is applied for the $\pi N$--T--matrix. We expect
this to be appropriate for the s--waves. The effect of this treatment for
the higher partial waves in the $\pi$N--system needs further study. 
We come back to this point when we discuss our results.

Results for total cross sections in all experimentally accessible channels 
are shown in Fig. \ref{total} as a function of $\eta$, 
the maximum momentum of the produced pions in units of the pion mass. 
Evidently the predictions
of the model are in good overall agreement with the data over a wide
energy range. We emphasize that the results for the 
channels $pp \rightarrow pn\pi^+$ and $pp\rightarrow d\pi^+$ 
do not involve any adjustable parameters and are, therefore,
genuine predictions of our model. In the case of $pp \to pp\pi^0$, however,
the basic model (including direct production plus rescattering)
yields only about 60\% of the measured cross section. (Corresponding
results are indicated by the dash-dotted curve in Fig.~\ref{total}.) 
Here we have added contributions from the HME mechanism, Fig.~\ref{beitraege}c,
and fixed their "strength" so that we can reproduce the data in the
near-threshold region (cf. Ref.~\cite{pap2}). Specifically, we have
included contributions due to $\omega$ exchange using the
vertex parameters $g^2_{\omega NN}/4\pi$ = 10 and $\Lambda_{\omega NN}$ 
= 1.5 GeV (monopole form factor). We would like to emphasize, however, 
that we do not view our HME contribution as being due to a genuine 
process but rather as an effective parametrisation of  
short range mechanisms \cite{KMR} not considered explicitly.

Note that, compared to our earlier
work \cite{pap1,pap2}, now both time orderings of the rescattering
diagram (Fig. \ref{beitraege}b as well as \ref{beitraegeND}d and e)
are properly included (the one where the pion is emitted
off one nucleon and interacts with the other before emission as
well as the one where one nucleon emits two pions, one of which
is absorbed on the other nucleon). This leads to a considerable
enhancement of the rescattering contribution so that now no additional
contribution from the HME mechanism due to the $\sigma$ meson is needed.
The effect of the HME contributions on the 
reaction channels $pp \to d\pi^+$ and $pp \to pn\pi ^+$ is
negligible \cite{pap2,NIS2}, so that the corresponding results remain 
practically unchanged.  Therefore we do not show them separately
in Fig.~\ref{total}. 

Let us discuss the influence of the $\Delta$ resonance on the 
production cross sections close to threshold. Our model includes pion
production from the $\Delta$ directly (Fig.~\ref{beitraegeND} a--c) or via $\pi N$
rescattering (Fig.~\ref{beitraegeND} d,e). The latter clearly gives 
contributions to s--wave pion production. In this context we wish to point out
that now -- unlike the purely nucleonic case -- charge--exchange rescattering 
is possible even in the reaction $pp \rightarrow pp\pi^0$ 
via a $\Delta^{++}n$ intermediate state.
However, it is less known that also direct pion production from the 
$\Delta$ gives a non-zero contribution at threshold. This can be easily seen 
from the standard reduction of the $\pi N\Delta$ vertex 
starting from the Lagrangian 
\begin{equation} 
{\cal L}_{\pi N\Delta} = {f_{\pi N\Delta}\over m_\pi} \bar \psi 
\vec T \cdot \partial^\mu \vec \Phi \psi_\mu \ + \ h.c. \ ,
\end{equation} 
where $\psi$, $\Phi$ and $\psi_\mu$ denote the nucleon, $\pi$ and $\Delta$ field
operators, respectively,
which leads to the following expression:
\begin{equation} 
M_{fi} \propto \lbrack \vec S^\dagger \cdot \vec q - 
{\vec S^\dagger \cdot \vec p \over M_\Delta} (\omega_q - 
{\vec q \cdot \vec p \over M_\Delta + E^\Delta_p} ) \rbrack \ . 
\end{equation} 
Here $\vec p$ ($E^\Delta_p = \sqrt{M_\Delta^2 + \vec p^2}$)
is the momentum (energy) of the incoming $\Delta$, 
$\vec q$ ($\omega_q=\sqrt{m^2_\pi + q^2}$) the momentum (energy) of the 
produced pion and $\vec S$ the spin transition operator. 
Evidently, even for vanishing pion momentum $\vec q$, the term 
proportional to $\omega_q / M_\Delta$, which is the analog to 
the recoil term appearing in the $NN\pi$--vertex, survives. 

Results for the production cross sections without inclusion of 
the $\Delta$ isobar are indicated in Fig.~\ref{total} by the dashed lines.
These curves are obtained by setting the $\pi N\Delta $ coupling in the
production operator to zero. Obviously, for the reaction 
$pp \rightarrow pp\pi^0$ the contributions involving the $\Delta$ lead to a
decrease of the cross section in the near threshold region. This reduction 
(by about 20~\%) is entirely due to the direct production mechanism; the 
contribution from rescattering off the $\Delta$ is negligibly small. At first
this is very surprising, especially because -- as was mentioned before --  
the $\Delta^{++}$ intermediate state allows also rescattering in the dominant 
$\pi N$ isovector channel. However, a detailed inspection of our results
reveals that the rescattering contribution is only small because
of a strong cancellation between the diagrams Fig. \ref{beitraegeND}d and e. 
Individually their magnitudes are quite significant.

Also for the reaction $pp \rightarrow d\pi^+$ we find that the direct 
production
is the dominant contribution among the ones involving the $\Delta$. It
increases the production cross section in the threshold region by about
30 \%. Thus, its contribution is partly responsible for the observed
overestimation of the $d\pi^+$ cross section close
to threshold. Still we should stress that the contributions from the
$\Delta$ that we get in our model are moderate as compared to the
ones reported by Niskanen \cite{NIS2}. In his case the inclusion of the
$\Delta$ leads to an increase of the $d\pi^+$ cross section close to threshold
by almost a factor of 3. We believe that this difference is due
a different treatment of the three particle propagator in the rescattering 
contribution involving the $\Delta$-resonance. Niskanen fixes 
the energy of the exchanged pion by putting it on--shell \cite{NIS4}. 
This choice maximizes the contribution of the $\Delta$ \cite{Mosb}. 

Analyzing powers for the reactions
$pp\rightarrow d\pi^+$, $pp \rightarrow pn\pi^+$ and $pp \rightarrow pp\pi^0$
at some selected energies are shown in Fig.~\ref{aypipl}. Evidently the 
data for this polarization observable are nicely reproduced by our model. 
This means that the model predicts the correct onset of higher
partial waves, especially of p--waves, and also the correct ratio of
p--waves to s--waves. The dashed curve in Fig.~\ref{aypipl} shows the 
results without contributions involving the $\Delta$ isobar. It is clear 
that the inclusion of the $\Delta$ isobar is essential for reproducing the 
data. It plays an important role even for 
energies very close to threshold ($\eta \leq 0.25$).  

Double polarization observables, namely the spin--dependent total cross section 
$\Delta \sigma_T/\sigma_{tot}=-(A_{xx}+A_{yy})$ and
the spin correlation coefficient $A_{xx}-A_{yy}$, have recently been
measured at IUCF for energies $\eta \geq 0.56$ \cite{IU4}.
In Fig.~\ref{axxayy} we present the predictions of our model for these 
observables.  Obviously the description of these data is not that good. 
Especially in case of $A_{xx}-A_{yy}$ our model
overestimates the data by a factor of about two. 
Since the result for $A_{xx}-A_{yy}$ without the $\Delta$ (dashed line)
looks much better one could get the impression that our treatment of the 
$\Delta$-resonance is incorrect. However, this observable is also very
sensitive to the non-resonant part of $\pi N$ p-wave rescattering. 
To illustrate this we show in Fig.~\ref{axxayy} a calculation,
where all contributions from $\pi N$ p-wave rescattering 
are switched off (dotted line). Also in this case
the description of the $A_{xx}-A_{yy}$ data clearly improves
whereas the results for $\Delta \sigma_T/\sigma_{tot}$ and $A_y$ 
remain practically unchanged. Thus, it seems that
$A_{xx}-A_{yy}$ is a particularly interesting observable for learning
more about the relation between the resonant and non-resonant contributions
from $\pi N$ p-wave rescattering. 

With regard to our model calculation we have already mentioned that 
we use an approximate description for the boost of the $\pi N$ T--matrix.
It is possible that this approximation leads to an overestimation of 
the contributions from $\pi N$ p-wave rescattering. 
Further studies in this direction are required.


In summary, we have presented a model for pion production in nucleon--nucleon 
collisions where the production operator 
is derived in a framework
consistent with the interaction potentials that are used for generating
the amplitudes in the elementary ($NN$ and $\pi N$) processes. The
$NN$ interaction includes explicit coupling to the $N\Delta$ channel
so that a consistent evaluation of pion production from $NN$ and
$N\Delta$ states can be done. The $\pi N$ amplitude is taken from 
a meson-exchange model of $\pi N$ scattering 
developed by the J\"ulich group. A good overall
description of cross section data for the reaction channels
$pp \rightarrow pp\pi^0$, $pp \rightarrow pn\pi^+$,
$pp \rightarrow d\pi^+$ and $pn \rightarrow pp\pi^-$ from the threshold up to the $\Delta$
resonance region is achieved. Quantitative
agreement, not only with integrated cross sections but also with analyzing
powers, is found over a wide energy range. Thereby the
inclusion of $\pi N$ rescattering as well as of the $\Delta$ degree of 
freedom plays an important role. We also demonstrated that polarization
observables are a powerful tool to investigate details of the dynamics of 
the process.

{\bf Acknowledgments}

We thank J. Durso for a careful reading of the manuscript. We also thank
Ulf-G. Mei\ss ner clarifying comments on various topics. One of
the authors (C.H.) is grateful for the financial
support by the COSY FFE--Project No. 41324880.

\begin{figure}[p]
\vspace{3cm}
\includegraphics{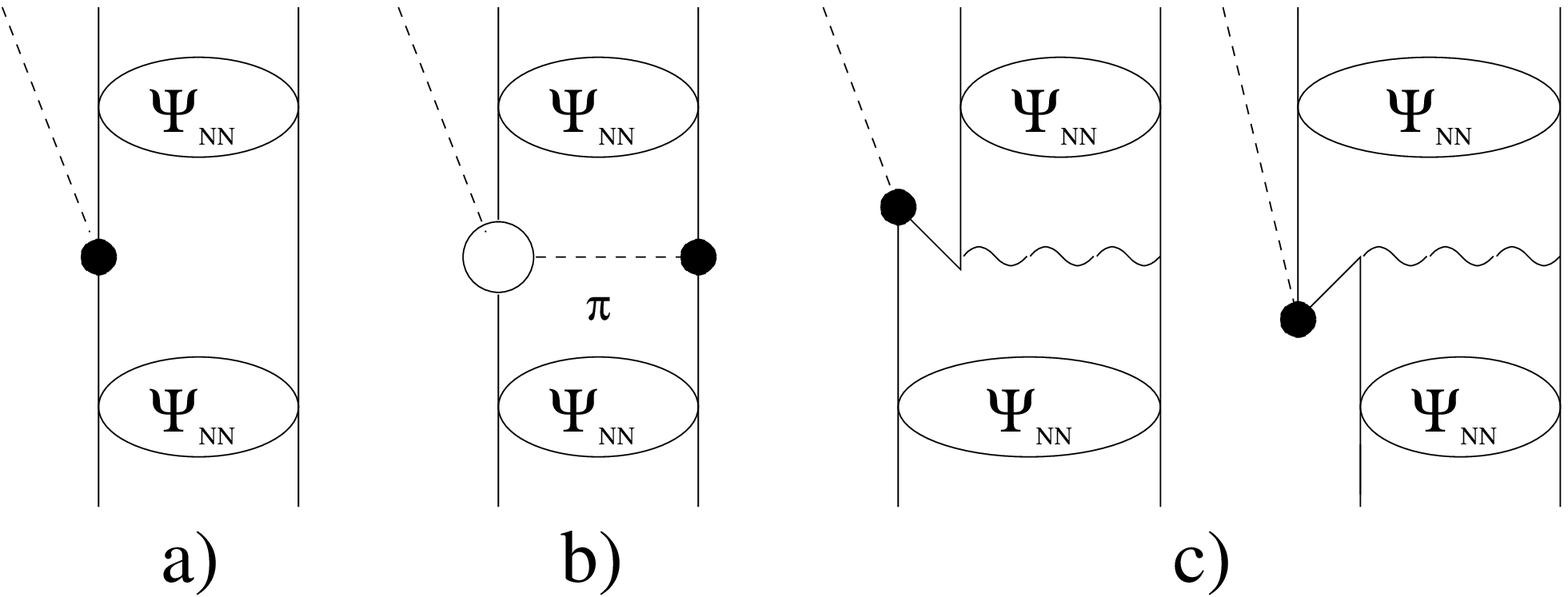}
\caption{\it{Pion production diagrams taken into account in our model
-- nucleonic contributions.}}
\label{beitraege}
\end{figure}

\begin{figure}[p]
\vspace{9cm}
\includegraphics{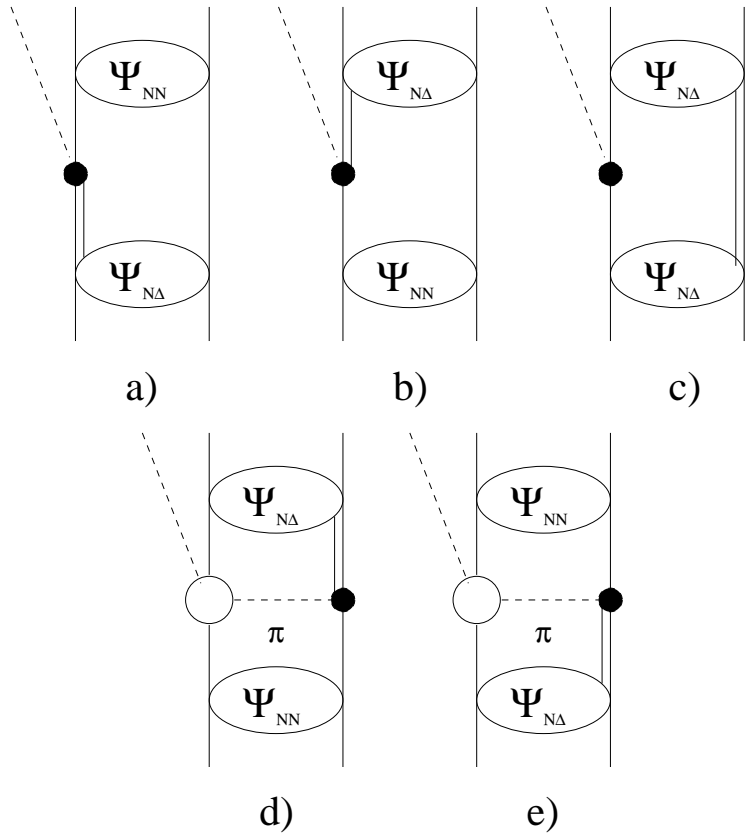}
\caption{\it{Pion production diagrams taken into account in our model
-- contributions involving the $\Delta$-resonance.}}
\label{beitraegeND}
\end{figure}

\begin{figure}[p]
\vspace{14cm}
\includegraphics{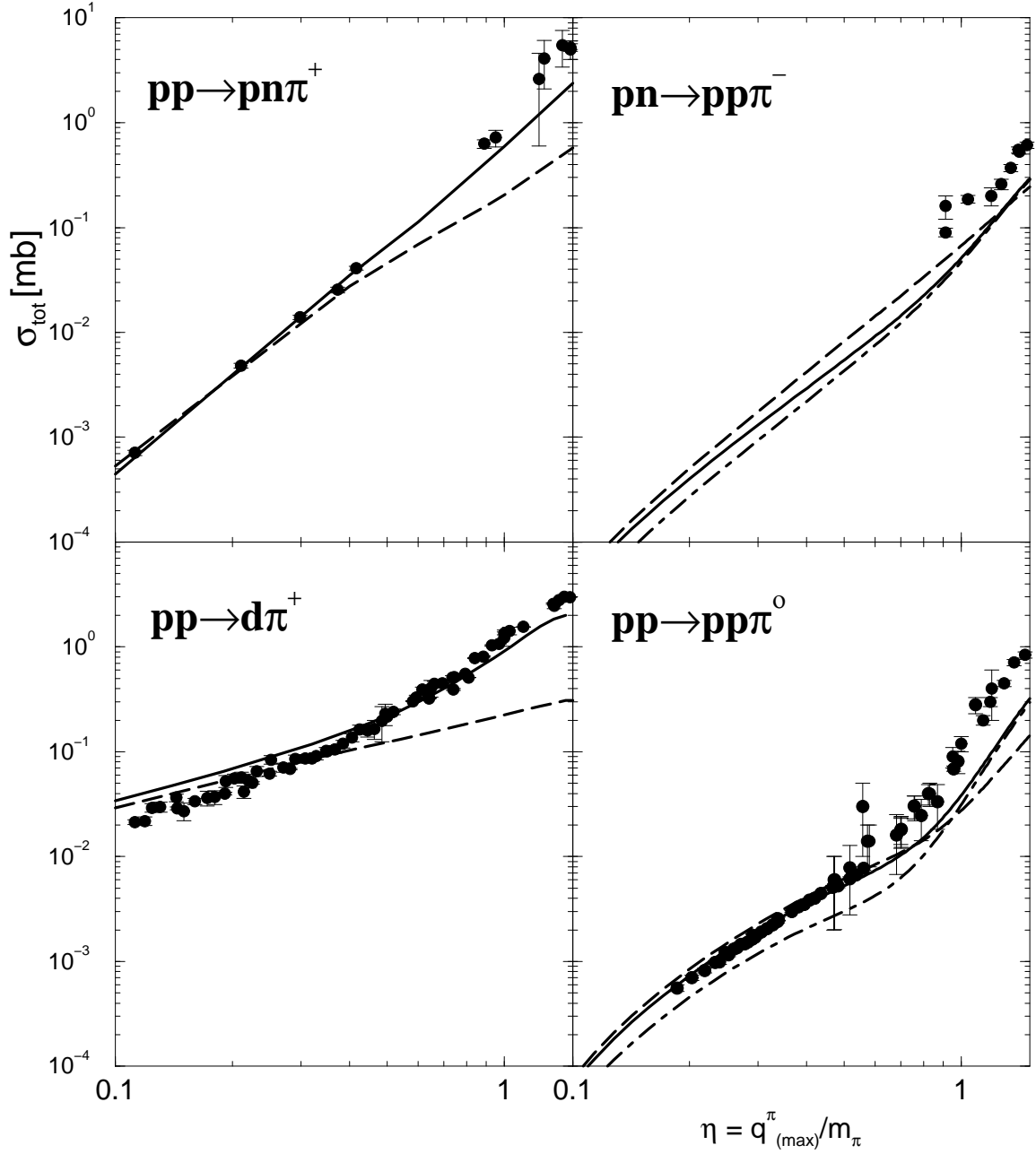}
\caption{\it{Total cross section for $NN\rightarrow NN\pi$ in the different
charge channels. The dash-dotted line shows the result of the
coherent sum of the direct production and the rescattering. For
the solid line heavy-meson-exchange contributions are added as described 
in the text. The dashed line shows the result without contributions 
involving the $\Delta$ isobar.}}
\label{total}
\end{figure}

\begin{figure}[h]
\vspace{20cm}
\includegraphics{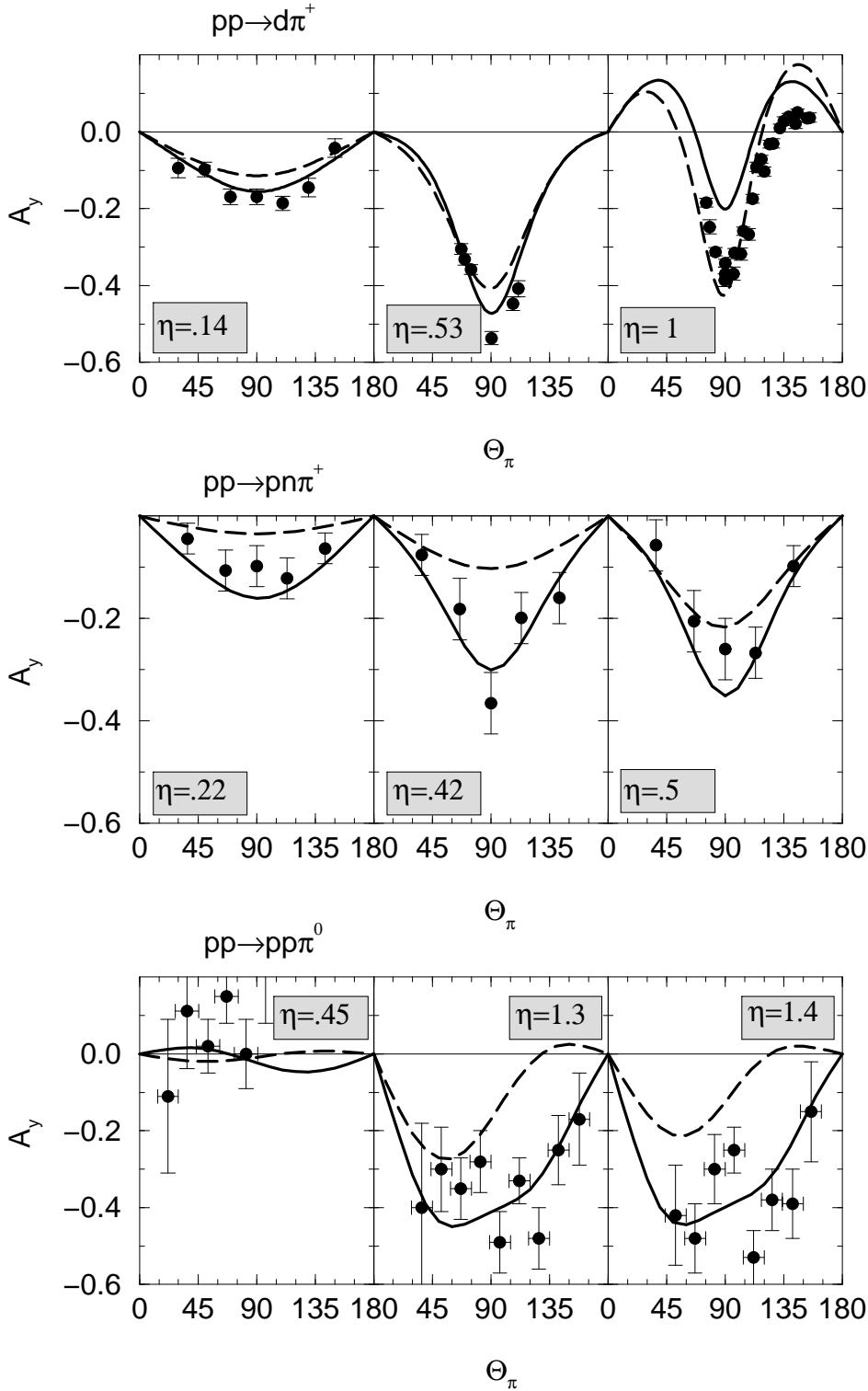}
\caption{\it{The analyzing power $A_y$. The solid line is the result of the 
full model, the dashed line is the result without the $\Delta$--isobar. 
Upper panel: $pp \to d\pi ^+$ at $T_{Lab} =$ 290.7, 330 and 425 MeV,  
respectively. Experimental data are from Refs. \protect{\cite{dpi1,dpi11}}. 
Middle panel: $pp \to pn \pi ^+$ at $T_{Lab} =$ 300, 320 and 330 MeV, 
respectively. Experimental data are from Refs. \protect{\cite{Ha}}.
Lower panel: $pp \to pp \pi ^0$ at $T_{Lab} =$ 310, 480 and 530 MeV, 
respectively. Experimental data are from Ref. \protect{\cite{pppi0sat}}.
}}
\label{aypipl}
\end{figure}

\begin{figure}[t!]
\vspace{10cm}
\includegraphics{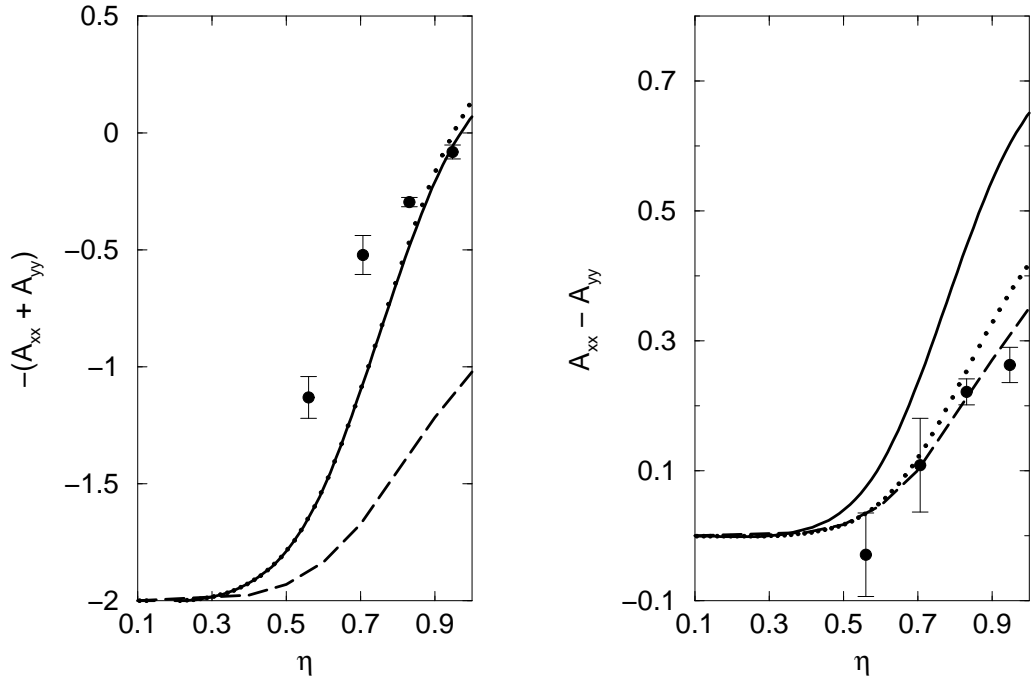}
\caption{\it{Predictions of our model for the spin--correlation functions
of the reaction $pp \rightarrow pp\pi^0$. The solid line is the result of the 
full model. The dashed line is the result without the $\Delta$--isobar,
whereas the dotted curve shows
the results, when the contributions from non-resonant $\pi N$ p-wave
rescattering are switched off. 
The data are from Ref. \protect{\cite{IU4}}.}}
\label{axxayy}
\end{figure}
 
\end{document}